\documentclass[a4paper,
               ]{jacow}
\usepackage{pdfpages,multirow,ragged2e} %
\makeatletter%
	\ifboolexpr{bool{xetex}}
	 {\renewcommand{\Gin@extensions}{.pdf,%
	                    .png,.jpg,.bmp,.pict,.tif,.psd,.mac,.sga,.tga,.gif,%
	                    .eps,.ps,%
	                    }}{}
\makeatother

\ifboolexpr{bool{xetex} or bool{luatex}} 
 {}                                      
 {\usepackage[utf8]{inputenc}}           

\usepackage[USenglish]{babel}

\ifboolexpr{bool{jacowbiblatex}}%
 {%
  \addbibresource{jacow-test.bib}
  \addbibresource{biblatex-examples.bib}
 }{}
\begin{document}

\title{Plasma Processing of FRIB Low-Beta Cryomodules using Higher-Order-Modes\thanks{This material is based upon work supported by the U.S. Department of Energy, Office of Science, Office of Nuclear Physics and High Energy Physics and used resources of the FRIB Operations, which is a DOE Office of Science User Facility under Award Numbers DE-SC0023633 and DE-SC0018362.  Additional support was provided by the MSU/FRIB Accelerator Science and Engineering Traineeship program.}}

\author{P. Tutt\thanks{tutt@frib.msu.edu}, W. Chang, K. Elliott, W. Hartung, S. Kim, K. Saito, T. Xu \\ Facility for Rare Isotope Beams, Michigan State University, East Lansing, MI, USA} 

\maketitle

\begin{abstract}
Improvement in SRF accelerator performance after in-tunnel plasma processing has been seen at SNS and CEBAF. Plasma processing development for FRIB quarter-wave and half-wave resonators (QWRs, HWRs) was initiated in 2020. Plasma processing on individual QWRs ($\beta = 0.085$) and HWRs ($\beta = 0.53$) has been found to significantly reduce field emission. A challenge for the FRIB cavities is the relatively weak fundamental power coupler (FPC) coupling strength (chosen for efficient continuous-wave acceleration), which produces a lot of mismatch during plasma processing at room temperature. For FRIB QWRs, driving the plasma with higher-order modes (HOMs) is beneficial to reduce the FPC mismatch and increase the plasma density. The first plasma processing trial on a spare FRIB QWR cryomodule was done in January 2024, with before-and-after bunker tests and subsequent installation into the linac tunnel.  The first in-tunnel plasma processing trial was completed in September 2025.  For both cryomodules, before-and-after cold tests showed a significant increase in the average accelerating gradient for field emission onset after plasma processing for some cavities.  In parallel with the cryomodule trials, the use of dual-drive plasma is being explored with the goal of improving the effectiveness of plasma processing.

\end{abstract}

\section{Introduction}
Superconducting radio frequency (SRF) cavities are a fundamental component of modern particle accelerators due to their extremely low dissipated power, enabling operation at high gradients and in continuous wave (CW) mode. SRF cavities are used in facilities around the world to accelerate electrons, protons, and ions for a variety of scientific purposes. While great care is taken with surface preparation and clean assembly to produce high-performance field emission (FE) free cavities, imperfect processes, subsequent assembly into cryomodules, and long-term accelerator operation provide opportunities for particulate contamination. As a result, FE may degrade the cavity performance. Surface treatments such as mechanical polishing, buffered chemical polishing (BCP), electropolishing (EP), and high-pressure rinsing (HPR) may remove contamination and improve the performance, but they cannot be performed in situ. For a cryomodule with degraded performance, disassembly is required to apply the aforementioned techniques. Disassembly and refurbishment of a cryomodule is time-consuming and costly. 

The Facility for Rare Isotope Beams (FRIB) is a user facility with 324 coaxial SRF cavities supporting nuclear physics experiments since 2022, with beam energies reaching up to 200 MeV per nucleon \cite{cavity}. As beam operations at FRIB continue, future FE degradation is of concern \cite{Ting talk @ ICABU}.

Plasma processing is an in-situ technique to improve the performance of SRF cavities and is being investigated at FRIB\@. Several accelerator facilities have reported success in improving the FE onset and thereby the usable accelerating gradient via plasma processing in spare cavities, offline cryomodules, and in-linac cryomodules \cite{SNS, SNS2, JLab, LCLS2, LCLS2-cryomodule, ANL}. We have conducted successful plasma processing trials on  spare FRIB cavities \cite{napac, srf23-coax} and on one offline FRIB cryomodule \cite{journal}. 

The first demonstration of plasma processing in the FRIB linac tunnel is reported in this paper. A higher-order mode (HOM) was used to ignite and sustain the plasma via the fundamental power coupler (FPC). The plasma systems and procedures will be discussed, along with the cryomodule performance in before-and-after cold tests. A parallel effort is underway to study plasma processing via dual-driven HOMs \cite{Tutt}; additional results from this effort will be presented.

\section{Plasma Processing: Basics}
Plasma processing was developed to remove hydrocarbon contamination on the inner surface of SRF cavities. A mixture of inert gas (Ar, Ne, or He) and oxygen is supplied to the cavity at room temperature. RF power is used to ignite a glow discharge inside the cavity using the FPC or an HOM coupler. Disassociation of the diatomic oxygen creates oxygen radicals that can react with hydrocarbon chains on the surface, producing volatile byproducts such as H$_2$O, CO$_2$, and CO\@. Active pumping on the cavity maintains the cavity pressure and allows for byproducts to be removed during processing. Monitoring of these byproducts by a residual gas analyzer (RGA) can provide information about the removal of contaminants. 

To enhance the density of oxygen radicals for more efficient cleaning, the electron number density ($n_{e}$) should be increased. This can be done by increasing the forward power ($P_f$) or increasing the drive frequency ($f_d$). Higher $P_f$ provides more energy to the plasma for production of more ionization. A higher $f_d$ is beneficial because the resonant frequency of the cavity shifts up due to presence of the plasma and associated decrease in the effective permittivity \cite{f-shift}. By shifting $f_d$ closer to the new resonance, more power is coupled into the cavity; however, this enhances the density and causes the resonance to shift further, as discussed in previous papers \cite{srf23-coax, LCLS2, JLab}. 

\section{Challenges and Development Efforts for FRIB Cavities}
The primary limitation for FRIB plasma processing is the significant mismatch of the fundamental mode when the cavity is warm, which raises concerns about breakdown within the coupler space. HOMs can be used to lessen this mismatch for FRIB QWRs \cite{napac}. Cold tests on both $\beta = 0.53$ HWRs and $\beta = 0.085$ QWRs before and after plasma processing showed an improvement in the field emission onset \cite{napac, srf23-coax}.

The procedures developed on QWRs were applied to a spare QWR cryomodule in January 2024 \cite{journal}.  Four out of eight cavities were plasma-processed using either the TEM $5\lambda/4$ mode ($\sim 400$ MHz), a dipole mode ($\sim 605$ MHz), or both.  Before-and-after cold tests showed an increase in the FE onset field for 3 out of 4 cavities.

\section{First In-Tunnel Plasma Processing Trial for the FRIB Linac}

Following the successful plasma processing trial with an offline QWR cryomodule, the first in-tunnel trial on a FRIB QWR cryomodule (SCM805/CB06) was done during the Summer 2025 maintenance period (August-September).
The $\beta = 0.085$ cryomodule showing the most field emission degradation was selected for this trial.  The preliminary results will be presented in this section, including the impact of plasma processing on field emission performance; a more detailed analysis of the plasma measurements, the process by-products, and impact on multipacting is planned.

\subsection{Plasma Processing System}

\begin{figure}[htp]
    \centering
    \includegraphics[width=\columnwidth]{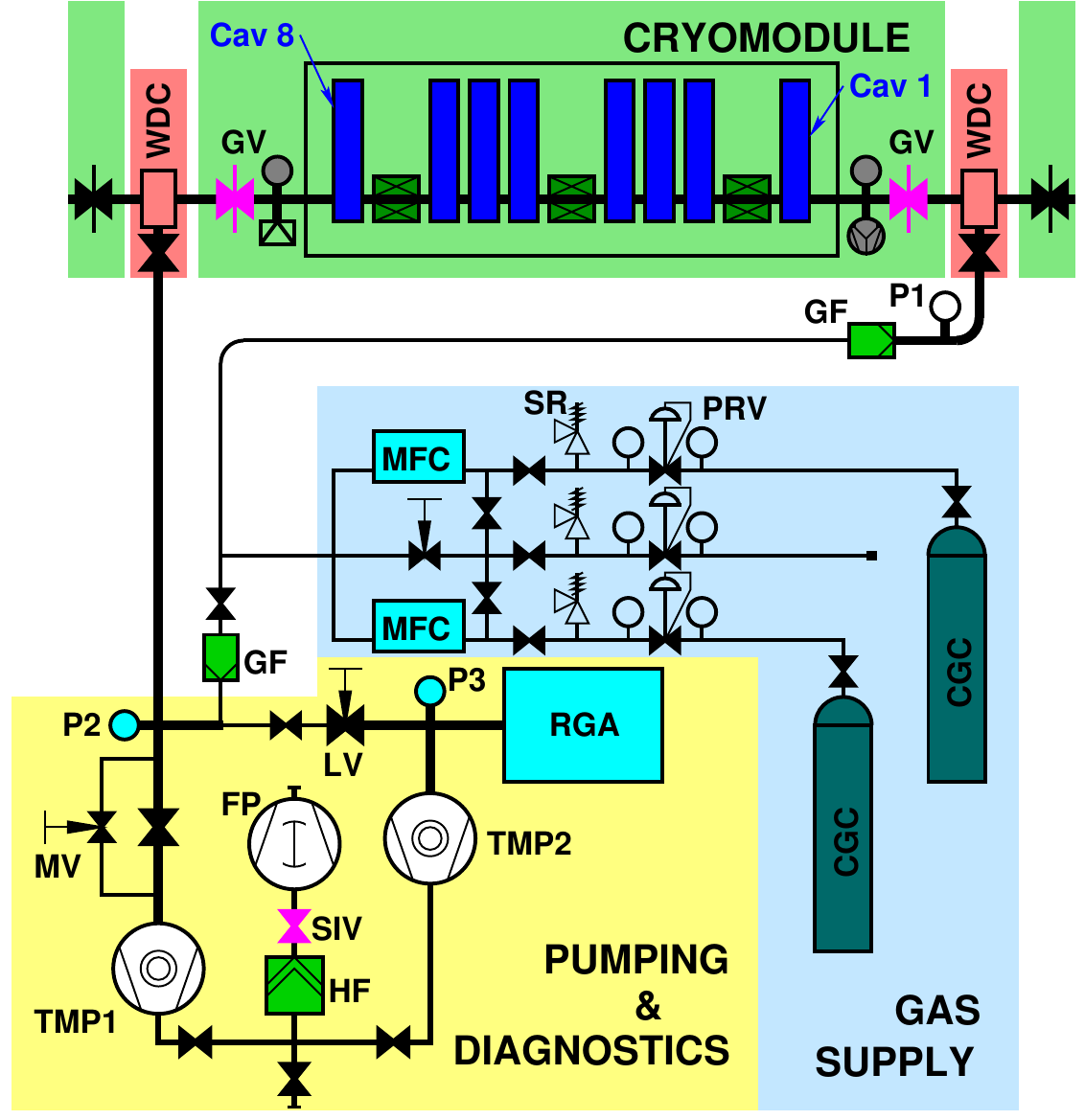}
    \caption{Diagram of the gas supply and pumping system. The gas flows from Cavity 1 to Cavity 8. CGC: compressed gas cylinder; PRV: pressure regulation valve; SR: safety relief valve; MFC: mass flow controller; GF: gas filter; P1, P2, P3: pressure sensors; WDC: warm diagnostics chamber; GV: gate valve; MV: metering valve; LV: leak valve. TMP1, TMP2: turbo-molecular pumps; FP: fore-line pump; HF: HEPA filter; SIV: sentry isolation valve. RGA: residual gas analyzer.  Magenta: valves set up to close if power is lost.\label{fig:schema}}
\end{figure}

\begin{figure}[htp]
    \centering
    \includegraphics[width=\columnwidth]{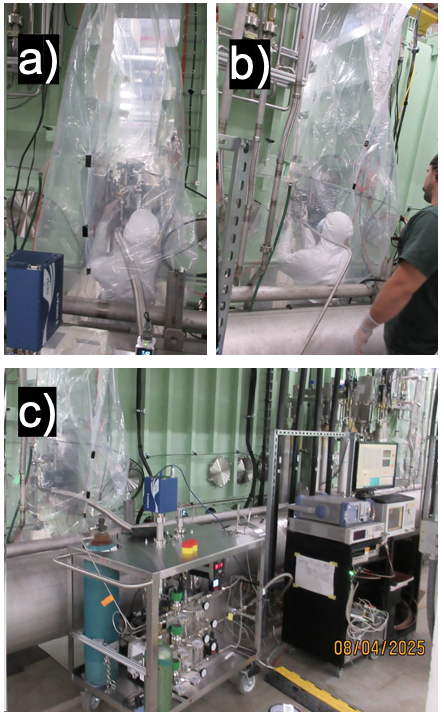}
    \caption{(a) Pumping connection to the downstream warm chamber. (b) gas supply connection to the upstream warm chamber. (c) gas supply/pumping cart, RF cart, and cryomodule.\label{fig:pics}}
\end{figure}

A gas supply and pumping cart for plasma processing of FRIB cryomodules was designed, fabricated, and used for both cryomodule trials, with some adjustments after the offline trial. Figure~\ref{fig:schema} shows a schematic of the cart and connections to the linac. Two gas streams (pure argon and an 80:20 mix of argon/oxygen) are combined upstream of the cryomodule. Mass flow controllers are used to obtain a 90:10 argon/oxygen ratio with the desired flow rate. Gas supply and pumping connections are made to warm diagnostic chambers upstream and downstream of the cryomodule.  As shown in Figure~\ref{fig:pics}, portable clean hoods are used to mitigate the risk of additional particulate contamination when making connections.  Pressure sensors are present on both the gas supply side and the pumping side. 

The plasma was ignited and sustained with a 100~W RF amplifier in continuous-wave mode via the FPC.  The RF system for driving and monitoring the plasma was similar to that used previously \cite{srf23-coax, journal}, with only minor adjustments.

\subsection{Plasma Processing Procedure}

Previous development at FRIB has explored the use of HOMs for plasma processing of QWRs and found success in improving FE onset \cite{napac, srf23-coax}. The most promising HOM so far is the TEM $5\lambda/4$ mode at $\sim 400$ MHz. CST \cite{CST} simulations show the electromagnetic field distribution has two zeros in the electric field, creating three high-electric-field lobes. The peak electric field occurs in the bottom lobe around the tip of inner conductor. This region overlaps well with the fundamental mode high electric field, which is the most important area to clean. CST simulations of the electromagnetic field distributions have been presented previously \cite{napac, Tutt}.

A gas mixture of argon and oxygen (90:10 ratio) was used at $\sim 60$ mTorr. The $P_f$ required for plasma ignition was between 20 and 25 Watts, and the processing was performed with $P_f \sim 12$ Watts. All cavities were driven with the same HOM, gas mixture, and approximate $P_f$ for processing. Some pressure differential was observed across the cryomodule. The processing was done in four rounds, as summarized in Table~\ref{T:rounds}.
The shift in the drive frequency ($\Delta f_{d}$, an indicator of the plasma density), cavities processed, number of iterations, and time per iteration are included. Based on previous experience with LCLS-II cryomodules \cite{paolo-talk}, all cavities received at least some plasma treatment even if they were FE-free. Round 1 and Round 4 done for  fast, half-density processing of all cavities. Round 2 targeted cavities with the lowest FE onset with full plasma density and multiple iterations. Round 4 targeted cavities with a higher FE onset, again with full plasma density and multiple iterations. In all rounds, the cavities were processed sequentially along the direction of the gas flow. 

\begin{table}[!hbt]
    \centering
    \caption{Plasma processing rounds.  The time values are per cavity and per iteration.\label{T:rounds}}
    \begin{tabular}{l|cccc} 
        \toprule
        \textbf{Round} & \textbf{1} & \textbf{2} & \textbf{3} & \textbf{4} \\
        \midrule
        Density & Half   & Full & Full & Half \\
        $\Delta f_d$ [MHz] & $\sim 0.4$   & $\sim 0.8$ & $\sim 0.8$ & $\sim 0.4$ \\
        Cavities & All  & 1, 4, 7   & 2, 3, 5 & All \\
        Iterations & 1  & 5 & 3  & 1 \\
        Time [Hr] & 0.5 & 1 & 1  & 0.5 \\
        \bottomrule
    \end{tabular}
\end{table}

\subsection{Pressure Excursion}
After the conclusion of plasma processing Round 4, the gas supply was closed to the cryomodule, and the pumps on the cart were used to restore the beam line vacuum. Prior to turning on the ion pumps, the beam line pressure was observed to quickly rise to $\sim 60$~mTorr and then begin to recover. Argon was the dominant peak seen on the RGA during this pressure excursion. We speculate that a power glitch caused the fore-pump and/or turbo-molecular pumps to stop pumping momentarily and then spontaneously recover. A power glitch was seen in other systems at FRIB and  across the wider Michigan State University campus where FRIB is located. The cryomodule gate valves were set up to  close in the event of power loss, but they remained open. Many other systems, including sensors and controls for plasma processing, appeared to be unaffected by the power disruption.

Possible back-streaming of contaminants during the pressure excursion was a concern. Two immediate steps were taken to protect the cryomodule: (i) the cryomodule was isolated from the gas supply and pumping cart and (ii) an additional sentry isolation valve was installed between the fore-line pump and turbo pumps (shown in magenta in Figure~\ref{fig:schema}).

\subsection{Additional Plasma Processing Rounds}

To mitigate the possible performance degradation due to back-streaming of contaminants, additional plasma processing rounds were performed. The ``bonus'' plasma processing rounds are summarized in Table~\ref{T:bonus}.  RGA signals during the bonus rounds did not show any evidence of additional hydrocarbons having been introduced by the pressure excursion.  Cavities closer to the pumping side of the cryomodule (Cavity 8 side) received more plasma processing time, since this is where contaminants from back-streaming would be more likely to be deposited.  The overall duration of the bonus rounds was set by the time available to finish maintenance work before the scheduled time for resumption of accelerator operations.

\begin{table}[!hbt]
    \centering
    \caption{Bonus plasma processing rounds.\label{T:bonus}}
    \begin{tabular}{l|cccc} 
        \toprule
        \textbf{Round} & \textbf{1} & \textbf{2} & \textbf{3} & \textbf{4} \\
        \midrule
        Density & Half  & Full & Full & Full \\
        $\Delta f_d$ [MHz] & $\sim 0.4$   &$\sim 0.8$& $\sim 0.8$ & $\sim 0.8$ \\
        Cavities & All  & All   & 5, 6, 7, 8 & 7, 8 \\
        Iterations & 1  & 1 & 1  & 1 \\
        Time [Hr] & 0.5 & 1 & 1  & 1 \\
        \bottomrule
    \end{tabular}
\end{table}

\subsection{Total Plasma Processing Time}

\begin{figure}
    \centering
    \includegraphics[width=\columnwidth]{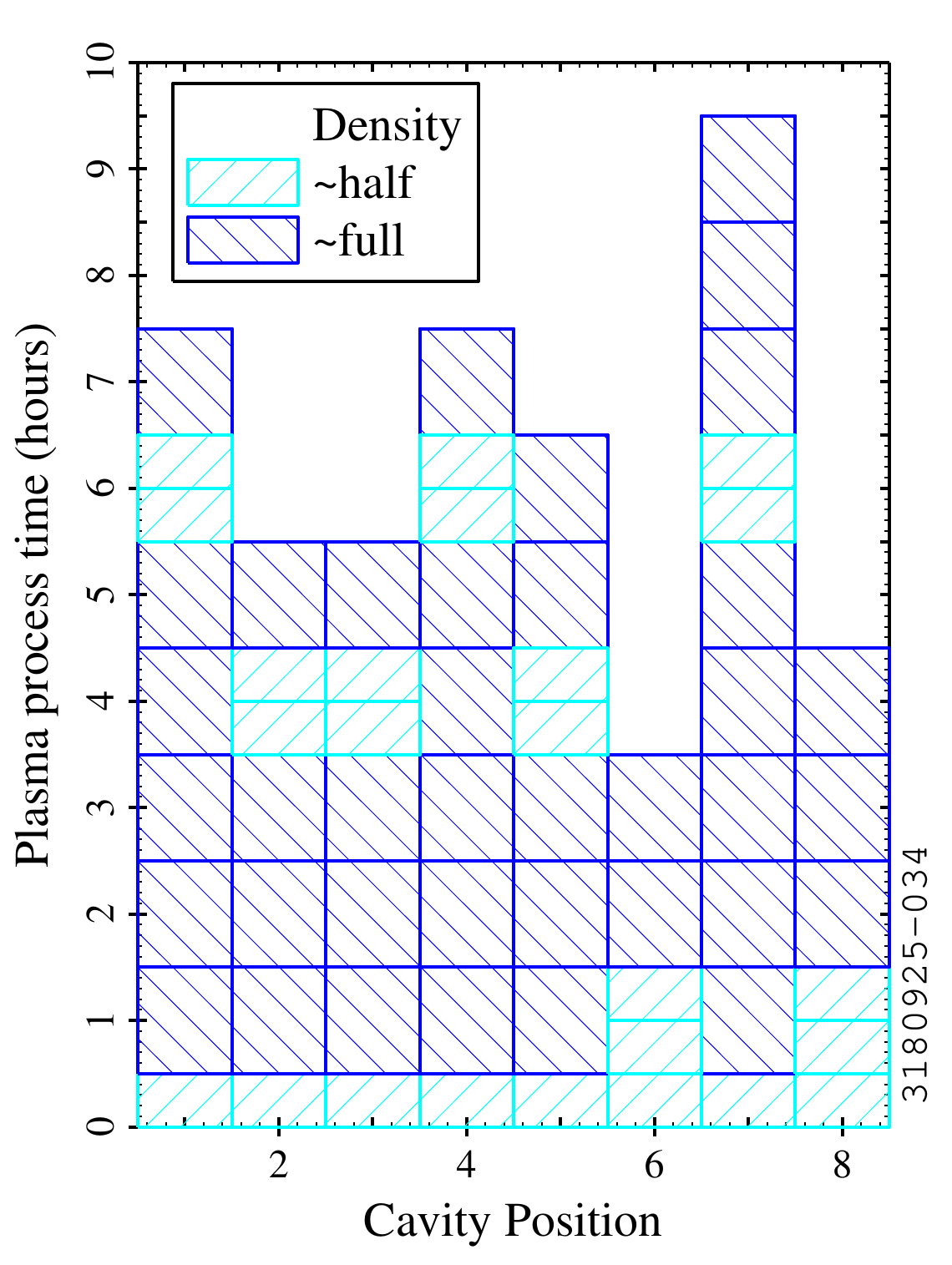}
    \caption{Plasma processing time for each cavity, including the originally-planned processing rounds and bonus rounds after the pressure excursion.\label{fig:barchart}}
\end{figure}

Figure~\ref{fig:barchart} shows the total plasma processing time for each cavity. Cavities 6 and 8 received the least plasma processing time since they were FE-free prior to plasma. Cavities 1, 4, and 7 received the most plasma processing time due to their early FE onset, while cavities 2, 3, and 5 received a more moderate plasma processing time due to their FE onset around the design field. Overall, Cavity 7 received the most processing, as it had a low FE onset and was close to the downstream end of the cryomdule.

\subsection{Before-and-After Field Emission Scans}

\begin{figure}[htp]
    \centering
\includegraphics[width=\columnwidth]{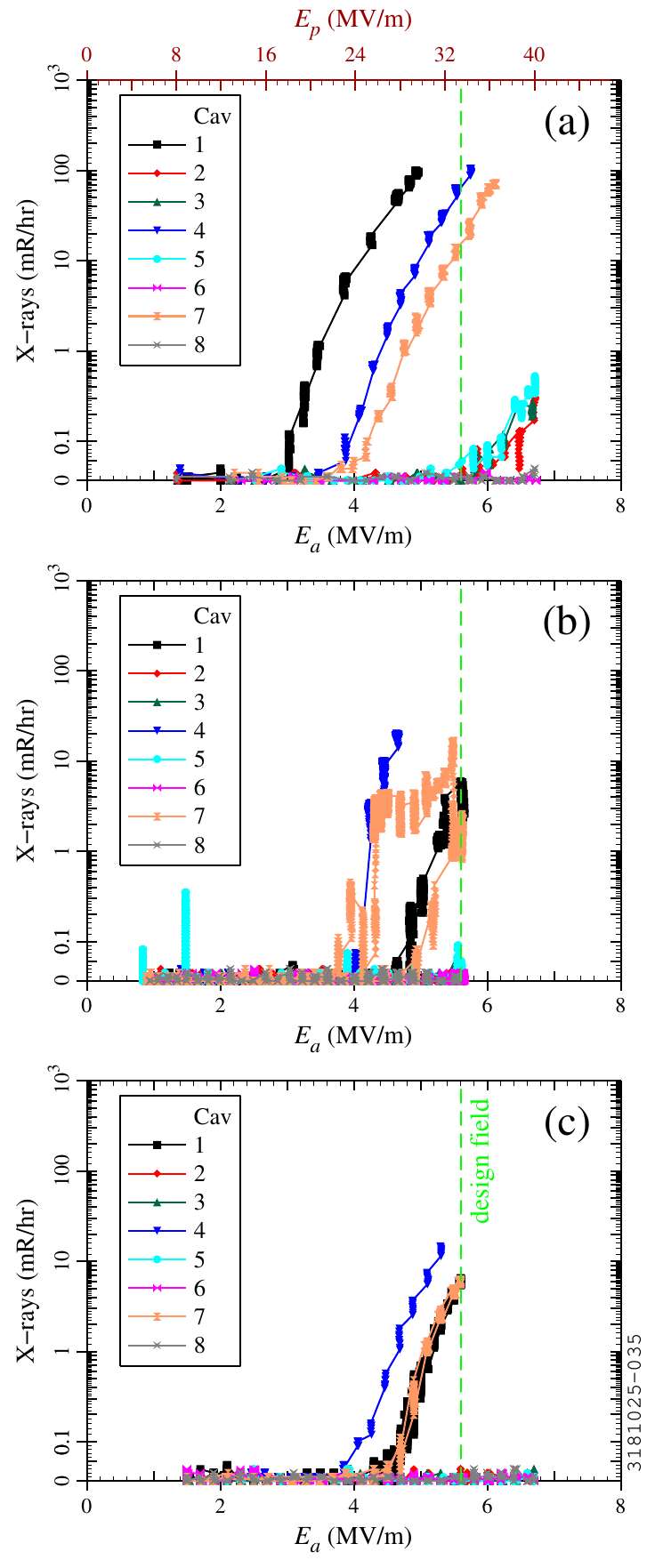}

    \caption{Field emission performance of the cryomodule (a) before the warm-up for plasma processing (July 2025); (b) after plasma processing (September 2025); (c) after additional CW/pulse conditioning on Cavity 7 and Cavity 4 (September 2025).  Upper axis: $E_p$ = peak surface electric field.  Dashed green line: design field.\label{fig:FEscans}}
\end{figure}

Figure~\ref{fig:FEscans}a shows the measured X-rays as a function of accelerating gradient prior to plasma processing. Three cavities showed FE onset below the design gradient ($E_a = 5.6$~MV/m, green dashed line); these cavities all experienced FE degradation between May 2022 and July 2025 and received 1 or more rounds of pulse conditioning for FE mitigation. Three cavities showed FE onset near or slightly above the design field. Two cavities were FE-free up to the FE-scan administrative limit of $E_a = 6.7$~MV/m.

Figure~\ref{fig:FEscans}b shows FE scans after plasma processing (administrative limit at $E_a = 5.6$~MV/m). The performance of the FE-free cavities (6 and 8) was preserved. Cavities 2, 3, and 5, which had FE onsets around the design field before plasma processing, did not show any FE up to the design field. Cavities with FE onsets below the design field before plasma processing (1, 4, and 7) showed mixed results: Cavity 1 had FE onset improvement from $\sim 3$~MV/m to $\sim 4.8$~MV/m. FE performance was similar for Cavity 4 and slightly worse for Cavity 7 after plasma processing. 

Pulse conditioning was done on Cavity 4 (after the initial post-plasma FE scan) and CW conditioning was done on Cavity 8 (during the initial post-plasma FE scan). An improvement in FE onset for Cavity 7 during the measurements can be seen Figure~\ref{fig:FEscans}b (orange hourglasses) due to the CW conditioning. After pulse conditioning of Cavity 4, the FE scans were repeated with an administrative limit of $E_a = 6.7$~MV/m, as shown in Figure~\ref{fig:FEscans}c. Cavities 2, 3, and 5 still showed no FE. Cavities 1 and 7 showed some improvement in FE onset; Cavity 4's FE onset remained similar.

\section{Dual-Tone \NoCaseChange{HOMs} for \NoCaseChange{QWRs}}

A clear disadvantage of HOM-driven plasma processing is the lack of uniformity in the plasma distribution.  The plasma distribution depends on the electromagnetic field distribution of the mode and which high-electric-field lobe of the mode ignites. Efforts are underway to study the usefulness of exciting two HOMs simultaneously to produce a more favorable plasma distribution. Studies of 
of the spatial plasma distribution with two HOMs were presented previously~\cite{Tutt}. We will expand on two other areas of interest in this section: reducing plasma ignition thresholds and moving the plasma from one lobe to another. 

\subsection{Reducing Ignition Thresholds}

Previous studies found two clear advantages to plasma processing at low gas pressure: (i) avoidance of coupler ignition and (ii) higher maximum drive frequency shift, allowing for higher plasma density \cite{LCLS2, hartung-hiat}. However, plasma ignition at lower pressures (in the range of 10's of mTorr, for example) typically requires higher forward power ($P_f$). Two HOMs of interest are the TEM $5 \lambda/4$ mode at $\sim 400$~MHz (used for cryomodule processing) and a quadrupole mode at $\sim 1$~GHz. Investigations of these modes with a neon and oxygen mixture (95:5 ratio) at $\sim 100$~mTorr showed ignition thresholds of $P_f \sim 35$~Watts for the quadrupole mode and $P_f \sim 50$~Watts for the TEM mode. Both modes ignite symmetrically in the cavity. Another HOM of interest due to its low ignition threshold is a 687 MHz dipole mode (ignition power $P_f \sim 7$~W). The latter mode ignites asymmetrically. Additional information about the plasma distributions for these modes can be found in Ref.~\cite{Tutt}.

Plasma ignition and extinction occur at different power levels such that more power is needed to ignite the plasma and less power is needed to sustain it \cite{srf23-coax, hosek-hiat}.
Modes with a low ignition threshold and an unfavorable plasma distribution can be used to ignite plasma, which can then be sustained using another drive mode: if a low-ignition-threshold mode is used to ignite plasma first, another high-ignition-threshold mode can be driven simultaneously with lower power; the first mode can then be turned off, causing the plasma to jump from one mode's distribution to the other. This has been demonstrated for the aforementioned dipole and quadrupole modes. The steps are
\begin{enumerate}
    \item Ignite the 687 MHz dipole mode with $P_f \sim 7$~W. 
    \item Drive the quadrupole mode at its perturbed frequency. 
    \item Ramp up the power for the quadrupole mode to a few W (well below the single-drive ignition threshold). 
    \item Turn off the dipole mode; check whether plasma is still driven by the quadrupole mode by monitoring both modes with a network analyzer. 
\end{enumerate}
Iterations in these steps may be needed to identify the quadrupole drive power range over which the plasma can be transferred.

\begin{figure}[htp]
    \centering
    \includegraphics[width=\columnwidth]{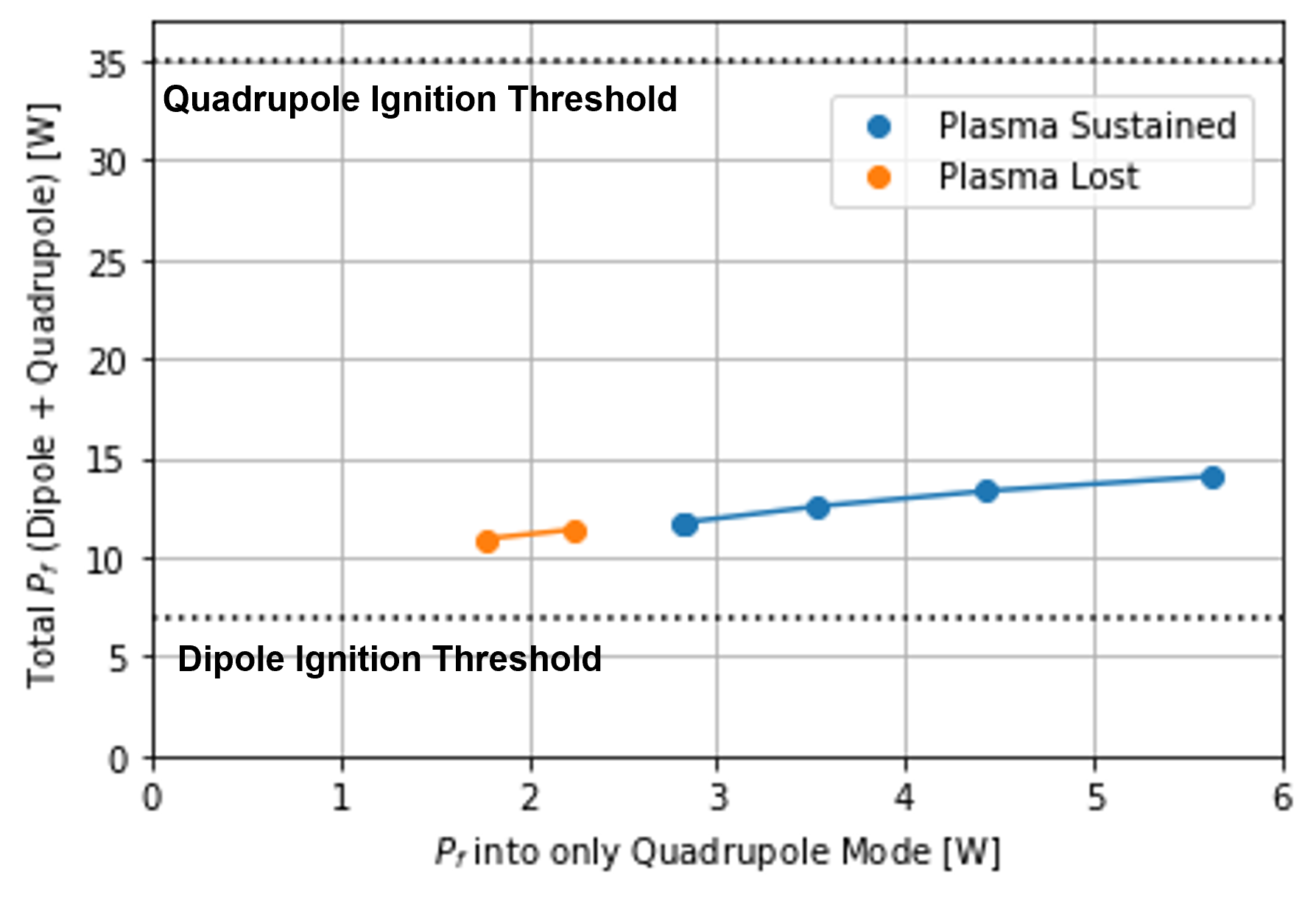}
    \caption{Reduction in the ignition threshold power for the $\sim 1$~GHz quadrupole mode with ignition-assist from a dipole mode at $\sim 690$~MHz.\label{fig:thresh}}
\end{figure}

Figure~\ref{fig:thresh} shows the total forward power into both modes as a function of the forward power into only the quadrupole mode for the gas mixture and pressure conditions mentioned earlier. The transition between plasma sustained in the quadrupole mode and plasma loss occurs at a total $P_f$ between 11.5 and 12.0~W and a quadrupole $P_f$ between 2.5 and 3.0~W\@. This ignition-assisted approach reduces the required power for quadrupole ignition by > 20 W. At this pressure, the better-uniformity plasma distribution of the quadrupole mode is achieved despite the asymmetry of the dipole mode.  

\subsection{Plasma Redistribution}

As reported previously, with a neon and oxygen mixture (95:5 ratio) at $\sim 100$ mTorr, dipole modes typically ignite in one high-electric-field lobe only, with a random ignition location. However, for the 687~MHz dipole mode, ignition is typically observed in one of the lobes at the bottom of the inner conductor, which is consistent with peak field locations from CST simulations \cite{Tutt}. Asymmetric ignition and subsequent plasma processing would likely lead to uneven cleaning of the cavity walls. Because of the lower ignition thresholds and higher plasma density that can be reached with dipole modes, mitigation of this asymmetric cleaning could be beneficial. One solution is to clean with the plasma in one lobe, move the plasma to the other lobe, and repeat the plasma cleaning for more complete surface coverage.

\begin{figure}[bp]
    \centering
    \includegraphics[width=\columnwidth]{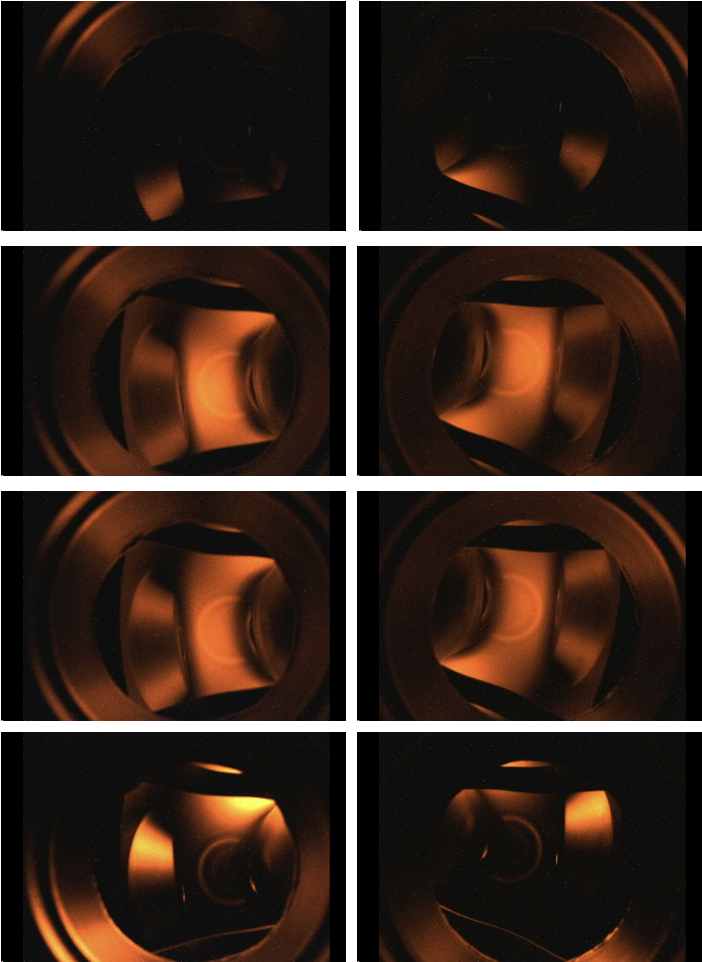}
    \caption{Transition of plasma between high-E-field lobes of the $\sim 690$~MHz dipole mode using the $\sim 1$ GHz quadrupole mode. Images are taken through viewports added to a QWR with a modified bottom flange, so the plasma may by viewed along the direction of the coaxial line. The two accelerating gaps span horizontally, the pickup-coupler is at the bottom of the images, and the FPC is at the top of the images. Top image: plasma ignition with dipole mode on pickup side. Middle top image: after driving and tuning the quadrupole mode. Bottom middle image: after retuning the dipole mode. Bottom image: after turning off the quadrupole mode.\label{fig:mix}}

\end{figure}

Figure~\ref{fig:mix} shows the variation in the plasma's spatial distribution while mixing the 687~MHz dipole mode and $\sim 1$~GHz quadruple mode. Images are taken using the same setup as described previously \cite{Tutt}. The top images show plasma ignition in the lobe on the pickup-coupler side of the cavity with $\sim 7$~W\@. After some tuning of the dipole mode drive frequency, the quadrupole mode was driven on its new perturbed frequency. After a sufficient increase in the quadrupole mode frequency, the plasma spontaneously jumps from the dipole distribution (top images) to the uniform quadrupole distribution (middle top images). By retuning the dipole $f_d$ up again, a smoother transition in the distribution is achieved, as seen in the middle bottom images: the light from the plasma is noticeably less bright in the accelerating gaps where the dipole mode has very little electric field. We interpret this to mean that the plasma distribution is shifting back to being localized in the quadrupole lobes that overlap well with the high-electric-field lobes of the dipole. If the quadrupole mode is turned off, the plasma returns to being localized in only one dipole lobe; however, the plasma is now in the opposite lobe, as shown in the bottom images. The difference in light intensity between the initial and final images is attributed to the shifting of the dipole drive frequency during the transfer of plasma between lobes. When the quadrupole mode is introduced, the dipole drive frequency shifts up. By retuning the dipole drive frequency and then turning the quadrupole off, the dipole resonant frequency is higher than in the initial image when plasma is first ignited. A higher shift in the resonant frequency indicates a higher plasma density, which is consistent with the observation of a brighter plasma.  The process is repeatable for ignition on the pickup side. In the absence of cameras and viewports, the location of the dipole-driven plasma can be determined from the measured dc currents on the FPC and pickup antennas.

\section{Conclusion}
In-situ plasma processing is being developed for FRIB cavities and cryomodules.  The first trial of plasma processing on a quarter-wave resonator cryomodule in the FRIB tunnel was completed in September 2025. After plasma processing, the field emission onset was higher for 5 out of 8 cavities, though 2 cavities required additional RF conditioning in CW or pulsed mode after cryomodule cool-down.  For the cavities with the lowest FE onsets, plasma processing and conditioning increased the FE onsets for 2 out of 3 cavities, but FE X-rays could still be seen at the design field.  Plasma processing of additional FRIB cryomodules is being considered for future maintenance periods.

Further optimization of plasma processing effectiveness and efficiency may be possible.  The use of dual-driven higher-order modes to improve the plasma distribution, reduce ignition thresholds, and improve surface coverage is being studied.  

Additional developmental work includes modeling of ignition and extinction thresholds for FRIB cavities \cite{hosek-hiat} using COMSOL \cite{COMSOL} and the application of COMSOL's plasma and RF modules to better understand the plasma dynamics and chemistry, following up on studies at Jefferson Lab for the CEBAF cavities \cite{jlab-sim, jlab-sim2}.

\section{Acknowledgments}

The authors wish to thank the participants in our informal plasma processing collaboration with IJCLab, Argonne, Fermilab, Brookhaven, and INFN-Milano for sharing their ideas, results, and suggestions. Colleagues at Jefferson Lab shared valuable advice about plasma modeling. We are grateful for the support from the cavity preparation, cryogenics, machining, RF, and vacuum teams at FRIB\@. Igor Nesterenko provided valuable guidance on plasma imaging.  We thank Zach Hosek, Sara Zeidan, and Jacob Brown for their helpful contributions to our plasma processing efforts.

%
%
\ifboolexpr{bool{jacowbiblatex}}%
	{\printbibliography}%
	{%

} 
%
%


\end{document}